\documentclass[aps,twocolumn,amssymb]{revtex4-2}
\usepackage{amssymb}

\usepackage{graphicx}
\usepackage{xcolor}
\usepackage{dcolumn}
\usepackage{bm}
\definecolor{linkcolor}{rgb}{0.0,0.3,0.5}
\usepackage[unicode, colorlinks=true, linkcolor=linkcolor, citecolor=linkcolor, 
filecolor=linkcolor,urlcolor=linkcolor, pdfusetitle]{hyperref}

\usepackage{makecell}
\usepackage{nccmath}

\def\d{{\textrm  d}}
\usepackage{stackengine}

\newcommand{\nn}{\nonumber}

\newcommand{\beq}{\begin{equation}}
\newcommand{\eeq}{\end{equation}}

\newcommand{\bea}{\begin{eqnarray}}
\newcommand{\eea}{\end{eqnarray}}

\def\IR{\mathbb{R}}

\begin{document}

\vspace*{-.1cm}
\title{Half the Schwarzschild Entropy From Strominger-Vafa Black Holes}

\author{Pierre Heidmann}
\email{heidmann.5@osu.edu}
\affiliation{Department of Physics and Center for Cosmology and AstroParticle Physics (CCAPP),
The Ohio State University, Columbus, OH 43210, USA}

\begin{abstract}

We construct a regular bound state of two extremal black holes in type IIB supergravity with opposite charges.  The bound state resembles the Schwarzschild black hole from the asymptotics to the photon ring and has the same mass and charges as the Schwarzschild black hole. On the one hand, we show that the bound-state entropy admits a microscopic description in terms of branes and antibranes in string theory. On the other hand, we demonstrate that the entropy approaches half of the Schwarzschild entropy as the mass becomes significantly larger than the Kaluza-Klein scales used to regularize the geometry.  This opens up the possibility of a brane/antibrane interpretation of the Schwarzschild entropy in string theory.

\end{abstract}

\maketitle

\section{Introduction}

Understanding the microscopic description of the Schwarzschild entropy stands as a key goal for a comprehensive theory of quantum gravity.  In string theory, the introduction of D-branes \cite{Polchinski:1995mt} — nonperturbative solitonic states carrying charges under Ramond-Ramond fields — provided a crucial framework to investigate the origin of the Bekenstein-Hawking entropy.  A significant breakthrough was achieved by Strominger and Vafa \cite{Strominger:1996sh,*Maldacena:1999bp}, who offered a microscopic description of a class of five-dimensional \emph{supersymmetric} black holes by studying a BPS system of branes in type IIB string theory. This involved matching the state counting within the brane system at weak coupling with the black hole entropy at strong coupling, resulting in an expression:
\begin{equation}
S = 2\pi \sqrt{N_1 N_5 N_P}, \nn
\end{equation}
where $N_1$, $N_5$, and $N_P$ are the numbers of D1 branes,  D5 branes and momentum charge within the BPS system.

However, extending such descriptions beyond supersymmetry encounters a significant challenge as the entropy of non-BPS states varies with the string coupling. Nevertheless, noteworthy microscopic descriptions have been derived for \emph{near-extremal} black holes by perturbatively introducing antibranes to a  brane system \cite{Callan:1996dv,*Horowitz:1996fn,*Breckenridge:1996sn} or by assuming the absence of interaction between branes and antibranes \cite{Horowitz:1996ay}.

Despite these efforts, unraveling the fundamental nature of black holes significantly far from extremality —  for which Schwarschild is the best representation — remains a challenge. While near-extremal analysis suggests a plausible brane/antibrane interpretation of the entropy,  the mechanism for which these degrees of freedom organize at weak and strong coupling remains unclear \cite{Danielsson:2001xe, *Emparan:2006it}.  

Another attempt to offer a microscopic description of the Schwarzschild black hole involves a correspondence between the black hole and a self-gravitating gas of hot string \cite{Horowitz:1996nw,*Horowitz:1997jc}. However, this correspondence does not leave the entropy invariant,  and the string star entropy fails to capture the (Mass)$^2$ growth of the Schwarzschild entropy.

This article develops a novel approach to derive a brane/antibrane origin of the Schwarzschild entropy. We construct neutral geometries directly within type IIB supergravity, arising from the backreaction of D1-D5 branes and $\overline{\text{D1}}$-$\overline{\text{D5}}$ antibranes with P and $\overline{\text{P}}$ momentum charges. Our focus is on a neutral bound state consisting of two extremal black holes as in \cite{Strominger:1996sh,*Maldacena:1999bp},  which we will call \emph{Strominger-Vafa black holes} for simplicity.  One is BPS and the other is anti-BPS and they carry opposite brane charges.  Unlike other constructions of black hole bound states in the literature \cite{NoraBreton1998,Manko_1993,*Alekseev:2007re,*Alekseev:2007gt,*Manko:2007hi,*Manko:2008gb,Emparan:2001bb}, the separation between the black holes is not singular.  The black holes form a bubble between them, counterbalancing their self-attraction and rendering the geometry asymptotic to four-dimensional Minkowski spacetime.

First, we demonstrate that the entropy of the bound state remarkably matches the microscopic degrees of freedom within the brane/antibrane system \cite{Strominger:1996sh,*Maldacena:1999bp}, despite being far from extremality:
\begin{equation}
S = 2\pi \left( \sqrt{N_1 N_5 N_P} + \sqrt{N_{\overline{1}}N_{\overline{5}}N_{\overline{P}}} \right)\,,\nn
\end{equation}
where $N_{\overline{1}}$, $N_{\overline{5}}$, and $N_{\overline{P}}$ represent the numbers of antibranes and antimomenta in the configuration.

Second, we show that gravitational constraints on the bound state make the quantized charges scale with the ADM mass, $\mathcal{M}$, such as $N_X, N_{\overline{X}} \propto \mathcal{M}^\frac{4}{3}$,  when the mass is much larger than the internal dimensions.  This ensures that the entropy scales like $\mathcal{M}^2$, analogous to the Schwarzschild entropy. More precisely, we show that the entropy is half the Schwarzschild entropy (in units $G_4=1$):
\begin{equation}
S  \,=\, 2\pi \mathcal{M}^2 \,=\, \frac{1}{2} \,S_\text{Schw}\,.\nn
\end{equation}

This represents the first regular construction in supergravity of a brane/antibrane configuration with a well-defined microscopic description of its phase space, nearly as vast as that of a Schwarzschild black hole.   Additionally, we analyze the spacetime structure of the bound state,  derive some thermodynamic properties, and initiate a discussion on explicit microstate constructions.  This final aspect involves leveraging the large families of microstate geometries of Strominger-Vafa black holes constructed to date  \cite{Warner:2019jll,*Bena:2022ldq,*Bena:2022rna}, thereby establishing a robust connection between the microstructure of supersymmetric black holes and that of neutral systems in string theory.

\section{Ernst formalism in type IIB}
\label{sec:typeIIB}

The static Ernst formalism has been recently generalized to supergravity theories in a series of works \cite{Heidmann:2021cms, Bah:2022pdn, Heidmann:2022zyd,*Bah:2021owp,*Bah:2021rki,Bah:2022yji, *Bah:2023ows}, laying the foundation for classical constructions of non-supersymmetric systems in string theory.

We work in type IIB supergravity on T$^6$ with electromagnetic flux corresponding to D1-D5 branes and P momentum (and potentially antibranes and antimomentum depending on the sign of the charges)  \cite{Heidmann:2021cms, Bah:2022pdn}.  The T$^6$ decomposes in two S$^1$ dimensions, $(\psi, y)$, with radii $(R_{\psi}, R_{y})$ where $y$ is the common direction of the D1 branes,  D5 branes and P momenta.  The remaining four compact dimensions form a T$^4$ wrapped by the D5 branes.  The external spacetime consists of a four-dimensional infinite spacetime, $(t, r, \theta, \phi)$.  The spacetime and brane structure is depicted in the Table \ref{tab1} below.

\setlength{\tabcolsep}{10pt} 
\renewcommand{\arraystretch}{1.5} 
\begin{table}[h]
 \centering
\begin{tabular}{|c|c|c|c|c|c|c|c|}
\hline
 & $t$ & $r$ & $\theta$ & $\phi$ & $\psi$ & $y$ & T$^4$\\
\hline
D1& $\leftrightarrow$ & $\bullet$ & $\bullet$ & $\bullet$ & $|$ & $\leftrightarrow$ &  $|$  \\
\hline
D5 & $\leftrightarrow$ & $\bullet$ & $\bullet$ & $\bullet$ & $|$ & $\leftrightarrow$ &  $\leftrightarrow$ \\
\hline
P & $\leftrightsquigarrow$ & $\bullet$ & $\bullet$ & $\bullet$ & $|$ & $\leftrightsquigarrow$ &  $|$ \\
\hline
\end{tabular}
\caption{\textit{The brane and spacetime configuration in type IIB supergravity.  The horizontal arrows represent the directions along which the branes are extended, the vertical lines represent smearing directions,  the curly arrows represent the components of the momentum waves,  and finally the dots represents potential local sources in the external space. } \label{tab1}}
\end{table}

The Einstein-Maxwell equations for static and axially symmetric solutions decompose in a set of decoupled electrostatic Ernst equations for the electromagnetic fields and also for internal deformations of the T$^6$ \cite{Heidmann:2021cms}.  This allows to generate regular topological structure in the spacetime from the deformation of the T$^6$ and support this structure with D1-D5-P flux.

In this article,  we consider solutions where a Kaluza-Klein (KK) bubble of size $\ell$ is nucleated along $\psi$, and the D1, D5, and P fluxes are set equal for simplicity. This results in a solution of minimal six-dimensional supergravity under KK reduction along the T$^4$ given by:
\begin{align}
ds^2 = & - \frac{dt^2}{Z^2}+(dy-T\,dt)^2 \nn \\
&\medmath{+Z\left[ \left(1-\frac{\ell}{r}\right)d\psi^2 + e^{3\nu} \left( \frac{dr^2}{1-\frac{\ell}{r}}+r^2 d\theta^2\right) +r^2 \sin^2\theta \d\phi^2 \right]}, \nn\\
 F =&\,  dH \wedge d\phi \wedge d\psi -dT\wedge dt \wedge dy \,. \label{eq:6dmetric}
\end{align}
The radial coordinate range is $r \geq \ell$.  The functions $(Z, T, H, \nu)$ depend on $(r, \theta)$,  are governed by static Ernst equations,  and characterize the D1-D5-P fluxes and their backreaction onto the geometry. In the absence of flux ($Z=1$ and $T=H=\nu=0$), the solution yields a KK bubble geometry,  so an Euclidean Schwarzschild solution where the $\psi$ circle degenerates smoothly at the origin of space $r=\ell$.

There is an extensive body of work dedicated to constructing and analyzing solutions of the static Ernst equations that correspond to bound states of charged black holes in four dimensions \cite{NoraBreton1998,Manko_1993,*Alekseev:2007re,*Alekseev:2007gt,*Manko:2007hi,*Manko:2008gb,Emparan:2001bb}. However, these solutions are unphysical due to the existence of string-like singularities, known as struts, which separate the black holes and prevent their collapse.  The present type IIB frame allows for the embedding of these solutions in higher dimensions and the resolution of struts using KK bubbles, akin to the method employed in \cite{Elvang:2002br,Bah:2021owp,Bah:2022yji,Astorino:2022fge}. This enables the construction of physical bound states of D1-D5-P black holes on topologically-nontrivial spacetime structures.

\section{Neutral bound state of two Strominger-Vafa black holes}
\label{sec:BS2}

We solve the $(Z, T, H, \nu)$ sector by two extremal points with opposite D1-D5-P charges located at both poles of the bubble, $r=\ell$ and $\theta=0$ and $\pi$ \cite{osti_4567857,Emparan:2001bb,NoraBreton1998,Bah:2022yji, *Bah:2023ows, HeidmannMehta}. The solution is given by \eqref{eq:6dmetric} with:
\begin{align}
Z & = 1+\frac{2M(2r+M-\ell)}{(2r-\ell)^2-\ell^2\cos^2\theta-M^2 \sin^2 \theta} , \nn\\
T & = \frac{2M \sqrt{\ell^2-M^2}\,\cos\theta}{(2r+M-\ell)^2-(\ell^2-M^2)\cos^2\theta} \,,\nn  \\
 H &=\frac{M \sqrt{\ell^2-M^2}(2r+M-\ell)\,\sin^2\theta}{(2r-\ell)^2-\ell^2\cos^2\theta-M^2 \sin^2 \theta}\,, \label{eq:2BHfields}  \\
  e^{\nu} & =1-\frac{M^2\sin^2\theta}{(2r-\ell)^2-\ell^2\cos^2\theta} ,  \nn
\end{align}
where $M$ can be interpreted as the ``brane mass,'' and the bubble size,  $\ell$,  also corresponds to the distance between both black holes.  The bound-state energy,  so the four-dimensional ADM mass after reduction along $(\psi,y)$,  is
\begin{equation}
\mathcal{M} = \frac{\ell+3M}{4}\,.
\label{eq:ADMMass}
\end{equation}
We work in units where the four-dimensional Newton constant, denoted as $G_4$, is set to 1.  Moreover,  $G_4$ is expressed in terms of the T$^4$ volume, $(2\pi)^4 V_4$, the string coupling  and length such as $G_4= \frac{ g_s^2 l_s^8}{8R_{\psi} R_{y} V_4}=1$.

The large-distance behavior of the electromagnetic potentials,  $(T, H)$,  indicates that the bound state is \emph{neutral} with equal D1-D5-P dipoles $\mathcal{J} = \frac{M}{2}\sqrt{\ell^2 - M^2}$:
\begin{equation}
T \sim \frac{\mathcal{J} \,\cos \theta}{r^2}\,,\qquad H \sim \frac{\mathcal{J} \sin^2 \theta}{r}\,.
\end{equation}

Note that the brane mass is constrained by the separation between both black holes $M\leq \ell$.  Beyond that limit,  the branes and antibranes are too massive for a given distance,  so that they would be within the ``Schwarzschild radius'' of the bound state,  rendering the solution \eqref{eq:2BHfields} physically invalid.

The solution is well behaved for $r>\ell$ and is asymptotic to $\IR^{1,3}\times$T$^2$. The spacetime ends at $r=\ell$,  as a KK bubble with localized branes and antibranes at its poles, as illustrated in Fig.\ref{fig:BS2BMPV}.

\vspace{-0.2cm}
\subsection{Internal structure}
\vspace{-0.2cm}

\begin{figure}
\begin{center}
\includegraphics[width= 0.4 \textwidth]{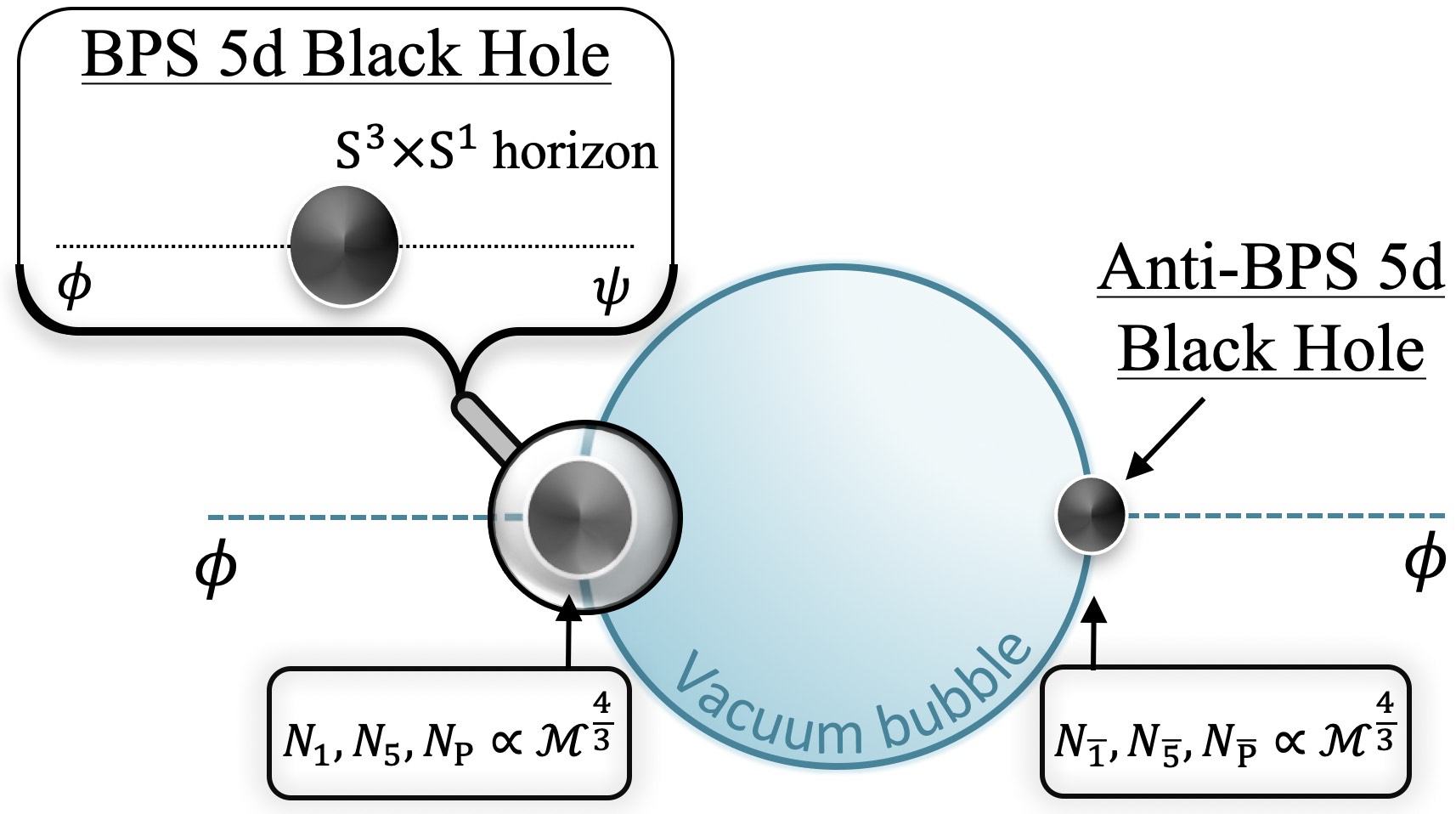}
\caption{\textit{The neutral bound state of two Strominger-Vafa black holes. The angles written indicate which circle smoothly degenerates on the symmetry axis.}}
\label{fig:BS2BMPV}
\end{center}
\vspace{-0.3cm}
\end{figure}

The S$^1$ parametrized by the $\psi$ coordinate degenerates at $r=\ell$ which ends the spacetime.  For $0 < \theta<\pi$, the geometry is most accurately described by the local coordinate $\bar{r}^2 \equiv 4(r-\ell)$ as $\bar{r} \to 0$.  The end-to-spacetime locus is smooth if the $(\bar{r},\psi)$ space defines an origin in $\IR^{2}$
\begin{equation}
ds_2^2= d\bar{r}^2 + \frac{\bar{r}^2}{R_\psi^2}\,d\psi^2,
\end{equation}
where $R_\psi$ is the radius of the S$^1$ defined by the periodicity $\psi=\psi+2\pi R_\psi$.  By expanding the solution \eqref{eq:6dmetric} with \eqref{eq:2BHfields},  we find that this requires to fix the bubble size, $\ell$,  in terms of the brane mass and $R_\psi$ such that
\begin{equation}
R_{\psi} = \frac{2(\ell^2-M^2)^\frac{3}{2}}{\ell^2}\,.
\label{eq:RegBu}
\end{equation}

Now, we shift our focus to the North pole of the bubble,  $r=\ell$ and $\theta=0$, by introducing local coordinates:
\begin{equation}
\medmath{\sqrt{r(r-\ell)} \sin \theta \equiv R \sin 2\tau \,,\quad \left( r-\frac{\ell}{2} \right) \cos \theta \equiv \frac{\ell}{2} + R \cos 2\tau.}
\end{equation}
As $R\to 0$, the local topology corresponds to an intricate S$^3$ fibration over AdS$_2\times$S$^1$, characterizing a BPS five-dimensional black hole, referred to as a Strominger-Vafa black hole \cite{Strominger:1996sh}:
\begin{equation}
ds^2 \sim d\widetilde{y}^2 + \frac{M(\ell+M)\Delta^2 }{2\ell^4 } \left[ ds(\text{AdS}_2)^2 +4 d\widetilde{\Omega}_3^2\right]
\end{equation}
where we have introduced $\Delta \equiv \ell^2-M^2 \cos^2\tau $,  $\widetilde{y}=y-\sqrt{\frac{\ell-M}{\ell+M}} \,t$,   $ds(\text{AdS}_2)^2 = \frac{dR^2}{R^2}-R^2 dt^2$,  and $d\widetilde{\Omega}_3^2$ is the line element of the three-sphere at the horizon parametrized by $(\tau,\phi,\psi)$. 

The single-center solution in \cite{Strominger:1996sh} has a round three-sphere at the horizon.  In our bound state,  the three-sphere at the black hole horizon is stretched, but remains regular \footnote{It is worth noting that the three-sphere is regular at its poles $\tau=0$ and $\pi/2$, as implied by the regularity constraint \eqref{eq:RegBu}.}: $$d\widetilde{\Omega}_3^2 =d\tau^2 + \frac{\ell^6}{\Delta^3} \left(\cos^2 \tau \,d\phi^2 +\frac{\sin^2\tau}{4\ell^2} \,d\psi^2 \right).$$

Moreover, the S$^1\times$S$^3$ horizon area is derivable, and the Bekenstein-Hawking entropy is given by:
\begin{equation}
S_\text{N} = \frac{\pi \left(M(\ell+M)\right)^\frac{3}{2}}{2^\frac{3}{2}\,\ell }.
\label{eq:EntropyNorth}
\end{equation}
The black hole is generated by the backreaction of D1 and D5 branes with P momentum charge.  The quantized charges,  that are derived from integrating the field strengths at the horizon,  are:
\begin{align}
N_1 &= \frac{ V_4 R_{\psi} M}{g_s l_s^6 }\,\sqrt{\frac{\ell+M}{\ell-M}} \,,\quad N_5 = \frac{R_{\psi} M}{g_s l_s^2} \,\sqrt{\frac{\ell+M}{\ell-M}}\,,  \nn \\
 N_P &= \frac{ V_4 R_{y}^2 \, R_{\psi} M}{g_s^2 l_s^8 }\,\sqrt{\frac{\ell+M}{\ell-M}} \,.
 \label{eq:QuantizedCharges}
\end{align}
Remarkably, the regularity at the bubble \eqref{eq:RegBu} forces the entropy to match  the microscopic degrees of freedom within the brane system, 
\begin{equation}
S_\text{N}=2\pi \sqrt{N_1 N_5 N_P}.\label{eq:StroVafaEntropy}
\end{equation}
Therefore, despite dealing with a non-supersymmetric solution that does not guarantee the entropy as a conserved quantity between strong-weak coupling, and despite having an extremal black hole for which the horizon is stretched by the bound state structure, the entropy still matches the microscopic value derived at weak coupling for single-center black holes in an asymptotically five-dimensional spacetime \cite{Strominger:1996sh}. \\

The South pole of the bubble,  $r=\ell$ and $\theta=\pi$,  leads to an identical geometry but with opposite charges. It corresponds to the anti-BPS partner of the Strominger-Vafa black hole at the North pole. It is produced by $\overline{\text{D1}}$-$\overline{\text{D5}}$ antibranes and $\overline{\text{P}}$ antimomenta, with the same quantized charges, $(N_{\overline{1}},N_{\overline{5}},N_{\overline{P}})=(N_1,N_5,N_P)$ (see Fig.\ref{fig:BS2BMPV}). It has therefore the same entropy $$S_\text{S} = 2\pi \sqrt{N_{\overline{1}}N_{\overline{5}}N_{\overline{P}}} = S_\text{N}.$$

We have successfully constructed a regular and neutral bound state involving two Strominger-Vafa black holes in type IIB on T$^4$. They are induced by branes and antibranes localized at distinct points in spacetime, generating a smooth bubble between them. The size of this bubble — the size of the bound state — is constrained by \eqref{eq:RegBu}, ensuring that the system is in equilibrium. The attractive force between the BPS and anti-BPS centers is counterbalanced by the topological pressure emanating from the bubble \cite{Elvang:2002br}. More surprisingly,  the degrees of freedom within the bound state align with the microscopic description of the extremal black holes.

A similar construction has been analyzed in \cite{Emparan:2001bb}, where neutral bound states of (near-)extremal black holes have been constructed in string theory and microscopically described in terms of branes and antibranes. However, this construction differs from the present one in crucial points: the solutions in \cite{Emparan:2001bb} do not have a KK bubble that regularizes the geometry between the black holes, resulting in a string singularity,  and the matching with the microscopic entropy in terms of branes and antibranes is only made when the black holes are widely separated, $\ell\gg M$. The presence of the strut makes the supergravity solution singular and nonphysical, and the black holes are four-dimensional extremal black holes with an S$^2$ horizon rather than Strominger-Vafa black holes with a three-sphere at their horizon.

Furthermore, as demonstrated in this section,  the regularization of the strut through a KK bubble introduces essential physics to the bound state given by the constraint \eqref{eq:RegBu}. First, it assigns a physical size to the bound state necessary for its equilibrium. Specifically, the separation of the black hole cannot be arbitrary; it closely approaches the brane mass when the KK scale is small,  $\ell-M = \mathcal{O}(R_\psi)$. Second, it establishes a connection between the Bekenstein-Hawking entropy, as derived in supergravity \eqref{eq:EntropyNorth},  and the microscopic entropy within the brane system \eqref{eq:StroVafaEntropy}.

\vspace{-0.2cm}
\subsection{Thermodynamics}
\vspace{-0.2cm}

The total entropy of the neutral bound state, denoted as $S \equiv S_\text{N} + S_\text{S}$, is
\begin{equation}
 S =  2\pi\left(\sqrt{N_{\overline{1}}N_{\overline{5}}N_{\overline{P}}}+\sqrt{N_1 N_5 N_P}  \right)  =   \frac{\pi \left(M(\ell+M)\right)^{\frac{3}{2}}}{\sqrt{2}\,\ell } .
 \label{eq:Entropy2BMPV}
\end{equation}
This microscopic description resembles the state counting of a non-extremal five-dimensional black hole assuming the branes and antibranes are not interacting \cite{Horowitz:1996ay},
\begin{equation}
S=2\pi(\sqrt{N_1}+\sqrt{N_{\overline{1}}})(\sqrt{N_5}+\sqrt{N_{\overline{5}}})(\sqrt{N_P}+\sqrt{N_{\overline{P}}}),
\label{eq:NearExEntropy}
\end{equation}  
but differs crucially in various aspects.

First, our bound state solution is net neutral, far from extremality, and corresponds to geometries that are asymptotic to a four-dimensional Minkowski spacetime.

Second, the brane/antibrane interaction in our bound state is captured in a nonperturbative manner, unlike the solution \eqref{eq:NearExEntropy}. In \eqref{eq:NearExEntropy}, the brane/antibrane interaction arises through cross-terms $N_{\text{I}}N_{\overline{\text{J}}}$ in the entropy,  while the values of $N_{\text{I}}$  and $N_{\overline{\text{J}}}$ have been obtained assuming the branes and antibranes do not interact.  Despite the absence of $N_{\text{I}}N_{\overline{\text{J}}}$ in the entropy formula of our bound state \eqref{eq:Entropy2BMPV},  this interaction is captured by the inherent dependence of the quantized charges in terms of the brane mass \eqref{eq:QuantizedCharges}. Indeed, when the black holes are widely separated, indicating the absence of interaction ($\ell \to \infty$), the quantized charges behave linearly as a function of the brane mass $M$ as expected. However, as the branes and antibranes come closer ($\ell \to M$) and interact more significantly, this linear relation undergoes a drastic change, and $N$ can become significantly larger than $M$. Consequently, a substantial portion of the brane energy (measured by the quantized charges) is used as the \emph{binding energy of the system}, and only a fraction contributes to the brane mass. Thus, the relationship between the quantized charges and the brane mass inherently captures the brane/antibrane interaction.  \\

Overall,  the bound state is determined by asymptotic quantities,  like the ADM mass, $\mathcal{M}$,  and the KK radius $R_\psi$.  Although we could not find a closed-form solution to invert the parametrization $(\ell,M)$ for $(\mathcal{M},R_\psi)$ using \eqref{eq:ADMMass} and \eqref{eq:RegBu}, we can derive the equivalence of the first law  by computing the variation of the entropy:
\begin{equation}
\frac{d\mathcal{M}}{M} = \frac{dS}{2 S} + \left(\frac{\pi^2 \ell^2}{ S^2 R_\psi} \right)^\frac{1}{3} \frac{dR_\psi}{8}\,.
\label{eq:FirstLaw}
\end{equation}
This is related to the generalization of the first law of black hole thermodynamics when considering a chain of black holes and bubbles \cite{Kastor:2008wd}. The first term on the right-hand side arises from the variation in the extremal black hole entropy, while the second arises from the variation in the bubble tension. 
At fixed $R_\psi$,  we have
\begin{equation}
\label{eq:FirstLaw}
d\mathcal{M} = \frac{\kappa}{2 \pi} \,dS \,,\qquad \kappa \equiv \frac{\pi M}{S}\,,
\end{equation}
where $\kappa$ is the \emph{effective} surface gravity of the bound state.  Thus,  we retrieve the first law of a Schwarzschild black hole,  but $\kappa$ depends on $M$ instead of $\mathcal{M}$.  The quantity $M$ being the brane mass, this implies that the entropy variation only arises from the branes and antibranes of the system and not the bubbling topology, as expected from the microscopic description \eqref{eq:Entropy2BMPV}.

\vspace{-0.2cm}
\subsection{Macroscopic limit}
\label{sec:Entrapment}
\vspace{-0.2cm}

The regularity \eqref{eq:RegBu} imposes a lower bound on the ADM mass and the size of the configuration:
\begin{equation}
\ell \geq \frac{R_\psi}{2}\,,\qquad \mathcal{M} \geq \frac{R_\psi}{8 }\,.
\end{equation}
The minimum occurs when there are no branes,  $M=0$,  and the geometry corresponds to a vacuum bubble, i.e., an Euclidean Schwarzschild geometry. More generally,  when $ \mathcal{M} \sim \tfrac{1}{8}R_{\psi}$, most of the mass arises from the bubble, and the branes and antibranes are just small perturbations on a vacuum bubble.  \\

In the macroscopic regime,  when the ADM mass is much larger than the KK scales $ \mathcal{M} \gg R_{\psi}$,  the energy in the bound state arises mainly from the branes and antibranes.  We obtain the following relations,  from \eqref{eq:RegBu} and \eqref{eq:ADMMass},  at leading order in $\epsilon \equiv \left( \tfrac{R_{\psi}}{2  \mathcal{M}} \right)^{2/3} \ll 1$,
\begin{equation}
\ell =  \mathcal{M} \left(1+\frac{3\epsilon}{8} \right)\,,\quad M =  \mathcal{M} \left(1-\frac{\epsilon}{8} \right)\,.
\label{eq:MacroRegime}
\end{equation}
Remarkably, this implies that the bound-state entropy becomes half of the Schwarzschild entropy,  $S_\text{Schw}=4\pi  \mathcal{M}^2$, and its effective surface gravity becomes twice the Schwarzschild value, $\kappa_\text{Schw}=(4 \mathcal{M})^{-1}$:
\begin{equation}
S = \frac{1}{2} \,S_\text{Schw}\,,\qquad \kappa = 2\kappa_\text{Schw}\,,
\end{equation}
with $\mathcal{O}(\epsilon)$ corrections. The equivalent first law \eqref{eq:FirstLaw} is then expressed as:
\begin{equation}
d\mathcal{M} =  \frac{\kappa_\text{Schw}}{\pi} \,dS + \frac{3  \mathcal{M}}{16}\,d\epsilon\,,
\end{equation}
where we have chosen $d\epsilon$ to capture  the variation in the bubble tension.

This is the main result of this letter. In type IIB supergravity, we have successfully constructed a \emph{neutral and regular} bound state of two extremal black holes, which exhibits \emph{similar thermodynamic properties to those of the Schwarzschild black hole}.  Moreover,  it has a clear microscopic origin in terms of the Strominger-Vafa fundamental description of five-dimensional BPS black holes  \cite{Strominger:1996sh}.

The scaling of the entropy as $\mathcal{M}^2$ is not trivial, considering its microscopic description \eqref{eq:Entropy2BMPV}. Typically, a BPS (or anti-BPS) system in string theory has  charges that scale with the energy,  $N \propto \mathcal{M}$. As such, one could expect $S \propto \mathcal{M}^\frac{3}{2}$. However, the charges for our bound state \eqref{eq:QuantizedCharges} behave,  at leading order in $\epsilon$,  as
\begin{equation}
(N_{\overline{1}},N_{\overline{5}},N_{\overline{P}}) = (N_1,N_5,N_P) \propto \left(\mathcal{M}^\frac{4}{3},\mathcal{M}^\frac{4}{3},\mathcal{M}^\frac{4}{3}  \right),
\label{eq:MacroQuantCharges}
\end{equation}
which is the correct power in $\mathcal{M}$ for the Strominger-Vafa entropy \eqref{eq:StroVafaEntropy} to scale as $\mathcal{M}^2$. Remarkably, this scaling is not something we had fine-tuned but emerged naturally from gravitational constraints. This unusual feature is due to the bound state nature of the solution. A significant part of the energy brought by the branes and momenta is used as binding energy. The ADM mass takes into account this fraction of energy lost and is therefore much less than what the quantized charges could indicate: $\mathcal{M} \propto N^\frac{3}{4}$. 

\vspace{-0.3cm}
\subsection{Spacetime structure}
\label{sec:GeoProp}
\vspace{-0.3cm}

In the macroscopic regime,  $\mathcal{M}\gg R_\psi$,  we have $\ell = M(1+\mathcal{O}(\epsilon))$ and the electromagnetic fields $(dH,dT)$ \eqref{eq:2BHfields} are of order $\epsilon$ everywhere except in proximity to the bubble at $r=\ell$. Moreover, the backreaction of the branes induces a significant redshift in that region, with $Z^{-1} \sim 1-\ell/r$. Consequently, the bound state exhibits a structure resembling a vacuum solution with a region of diverging redshift: a black geometry.

This behavior aligns with the notion of electromagnetic entrapment in gravity described in \cite{HeidmannMehta} for configurations of self-gravitating extremal charges.  In \cite{HeidmannMehta},  it has been shown that,  when a spacetime structure sourced by intense electromagnetic flux reaches a critical size associated to its ``Schwarzschild radius,'' the structure entraps its own electromagnetic field so that the solution looks like a neutral black hole up to the horizon scale.   

For our bound state, the geometries are indistinguishable from an embedding in higher dimensions of the $\delta=2$ Zipoy-Voorhees metric \cite{zipoy1966topology,*voorhees1970static,*stephani2009exact,Kodama:2003ch}.  Indeed,  for $r\gtrsim \mathcal{M}(1+\mathcal{O}(\epsilon^\frac{1}{4}))$, the electromagnetic field vanishes,  $F \sim 0$,  and the metric is
$ds_6^2 \sim ds_\text{ZV}^2 + d\psi^2 + dy^2$,
where $ds_\text{ZV}^2$ is the $\delta=2$ Zipoy-Voorhees metric
\begin{align}
ds_\text{ZV}^2 =& -\left(1-\frac{\mathcal{M}}{r}\right)^2 dt^2+ \frac{r^2\sin^2 \theta}{1-\frac{\mathcal{M}}{r}}d\phi^2 \label{eq:DisSchw} \\
&+\frac{\left(1-\frac{\mathcal{M}}{r}\right)^2}{\left(1-\frac{\mathcal{M}}{r}+\frac{\mathcal{M}^2 \sin^2\theta}{4r^2}\right)^{3}} \left(\dfrac{dr^2}{1-\frac{\mathcal{M}}{r}}+r^2 d\theta^2 \right). \nn
\end{align} 
This solution is an S$^2$ deformation on a Schwarzschild metric,  inducing a singular horizon where the S$^2$ gets flattened at its poles (refer to \cite{Kodama:2003ch,HeidmannMehta} for more details).  Despite this singularity, the horizon area remains finite and notably equals the Schwarzschild value, $16 \pi  \mathcal{M}^2$ \footnote{In \cite{HeidmannMehta},  the $\delta=2$ Zipoy-Voorhees metric has been called the distorted Schwarzschild black hole for these reasons.}.

In this context, the bound state replaces the singular horizon of a $\delta=2$ Zipoy-Voorhees solution with a regular and topologically-nontrivial structure in type IIB supergravity.  This resolution occurs in an infinitesimal region above the horizon,  and the size of this region is determined by the KK scales,  $\epsilon^\frac{1}{4} \mathcal{M}$.  The topology is supported by intense electromagnetic flux with a clear microscopic origin from branes and antibranes.  

\begin{figure}
\begin{center}
\includegraphics[width= 0.3 \textwidth]{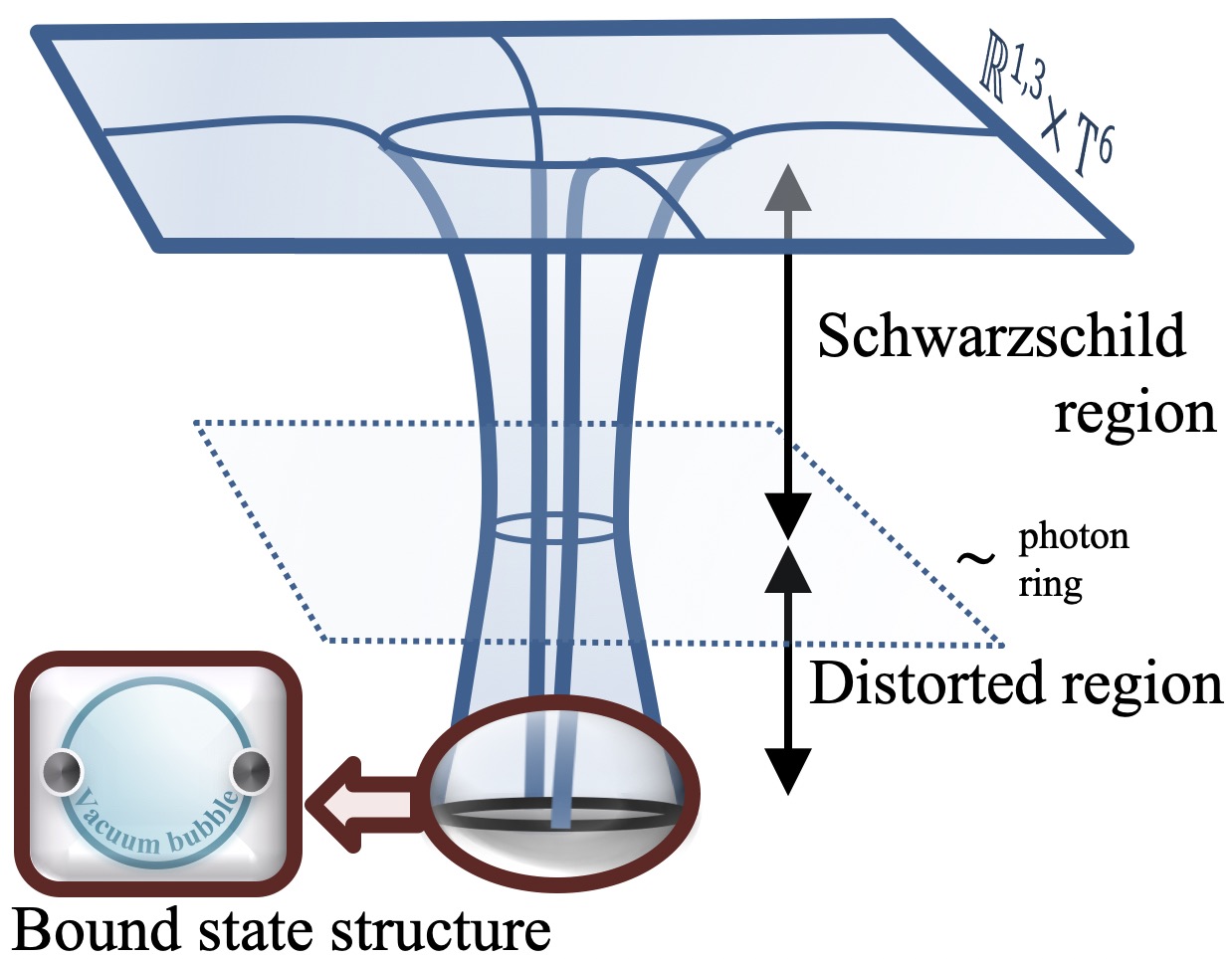}
\caption{\textit{Spacetime structure of the neutral bound state of two Strominger-Vafa black holes in type IIB supergravity.}}
\label{fig:Spacetime}
\end{center}
\end{figure}

Furthermore, the $\delta=2$ Zipoy-Voorhees geometry preserves numerous properties of the Schwarzschild black hole outside its singular horizon, including the two-sphere size, light ring characteristics, and gravitational signature \cite{Kodama:2003ch,HeidmannMehta}. The main deviations arise in the region between the Schwarzschild light ring and the horizon, specifically when the S$^2$ radius is smaller than $3\mathcal{M}$. 

Consequently, the spacetime generated by a bound state of two Strominger-Vafa black holes can be divided into three zones (see Fig.\ref{fig:Spacetime}): a Schwarzschild region from the asymptotics to approximately the light ring, a distorted region where the S$^2$ deformation intensifies and stretches the sphere, and the bound-state region where the structure of branes and antibranes starts to emerge at the place of the horizon of the Zipoy-Voorhees solution.

\begin{figure*}
\begin{center}
\includegraphics[width=  0.7 \textwidth]{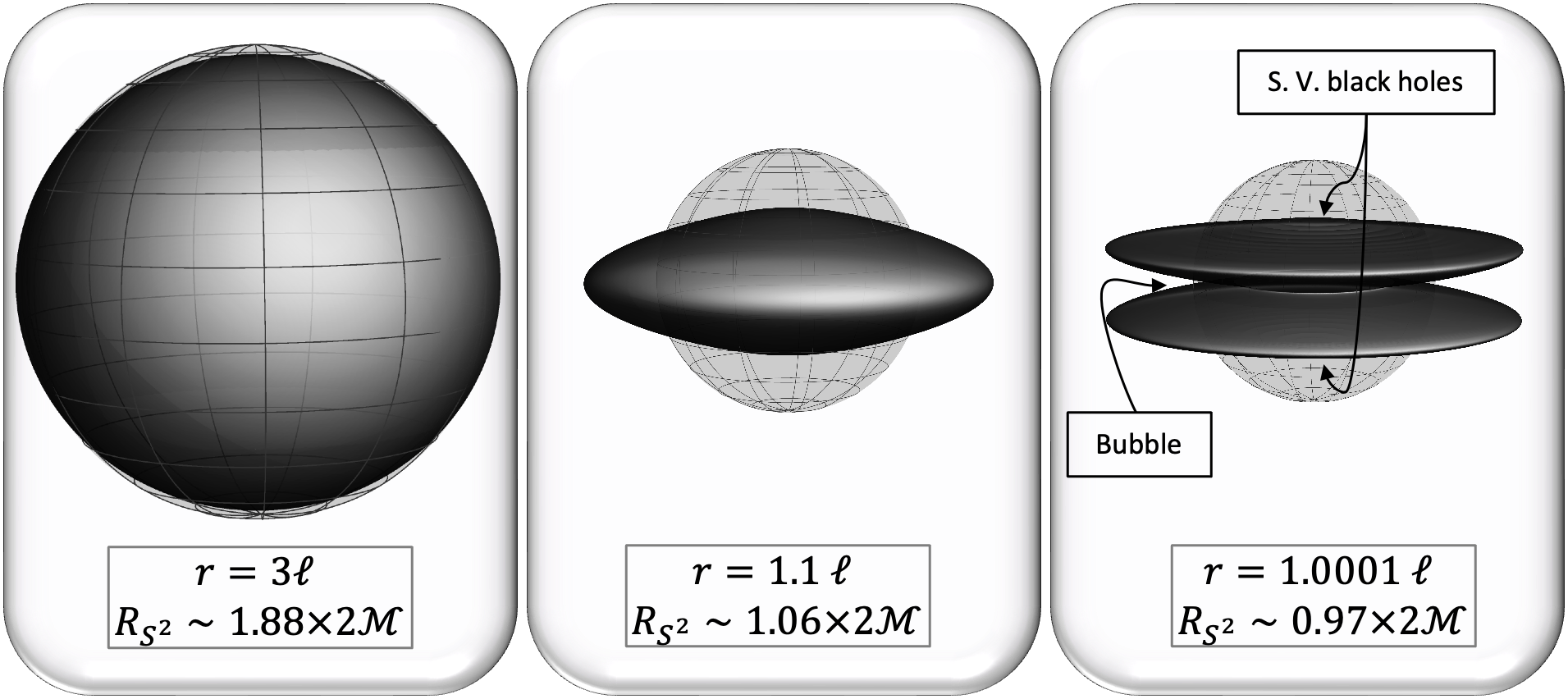}
\caption{\textit{Two-sphere surrounding the bound state at three distinct radial distances from the coordinate boundary $r=\ell$, where the Strominger-Vafa black holes and the bubbles are located. The two-sphere geometry is described by \eqref{eq:2Sphere}, with a faded round sphere of radius $R_{\text{S}^2}$ overlaying it to represent the equivalent Schwarzschild two-sphere. The value of $R_{\text{S}^2}$ is derived from the area \eqref{eq:overallradius}.  These illustrations are based on a bound state with $R_\psi/\mathcal{M}=10^{-1}$.   Moving from the left plot to the right, we observe the sphere in the region near the photon ring, followed by the distorted region, and finally, the vicinity of the bound state locus.}}
\label{fig:2Sphere}
\end{center}
\vspace{-0.3cm}
\end{figure*}

To demonstrate the proximity between the spacetimes produced by our bound state and a Schwarzschild black hole, we can provide a more detailed description of the two-sphere surrounding both geometries. In the case of Schwarzschild, the two-sphere is simply a round surface described by the line element $ ds(S^2)= r^2 (d\theta^2+\sin^2 \theta d\phi^2)$, with a radius $R_{\text{S}^2}=r$. For the bound states, obtained from the four-dimensional reduction of the metric \eqref{eq:6dmetric}, we have:
\begin{equation}
ds(S^2)= r^2 Z^\frac{3}{2} \sqrt{1-\frac{\ell}{r}} \left(e^{3\nu} d\theta^2 +\sin^2 \theta \,d\phi^2 \right)\,.
\end{equation}
To describe the two-sphere, we follow the approach outlined in \cite{Costa:2000kf}. We introduce a change of angular coordinate so that the line element becomes:
\begin{equation}
ds(S^2)= R(r,\widetilde{\theta})^2 \left(d\widetilde{\theta}^2 +\sin^2 \widetilde{\theta} \,d\phi^2 \right)\,,
\label{eq:2Sphere}
\end{equation}
where $R(r,\widetilde{\theta}) $ parametrizes a two-dimensional surface representing the two-sphere. Additionally, we define the average two-sphere radius $R_{\text{S}^2}(r)$ from the area:
\begin{equation}
 R_{\text{S}^2}(r)^2\, \equiv \,  \frac{1}{2} \,\int_{\theta=0}^\pi R(r,\widetilde{\theta})^2 \sin \widetilde{\theta}\, d\widetilde{\theta\,}.
\label{eq:overallradius}
\end{equation}
In Fig.\ref{fig:2Sphere}, we plot the two-sphere using $R(r,\widetilde{\theta}) $ for three representative $r$ coordinates,  and for a bound state with $R_\psi/\mathcal{M}=10^{-1}$, along with a round sphere of radius $R_{\text{S}^2}$ for comparison with a Schwarzschild black hole. The first plot depicts the region around the photon ring where $ R_{\text{S}^2}\sim 3\mathcal{M}$ and indicates a close resemblance to the Schwarzschild black hole's geometry.  This shows that for any distance $r\gtrsim 3\mathcal{M}$,  the spacetime is mainly indistinguishable from the Schwarzschild black hole as illustrated in Fig.\ref{fig:Spacetime}.  The middle plot shows the distorted region where the bound state closely resembles the $\delta=2$ Zipoy-Voorhees geometry, with the S$^2$ deformation amplifying. Finally, the last plot displays the region very close to $r=\ell$, where the regular inner structure of the bound state replaces the singular horizon of the Zipoy-Voorhees geometry. It is in this region that the two black holes and the bubble become visible, along with their electromagnetic fields. This plot clearly shows two highly stretched horizons due to the intense gravitational field between them.

In conclusion, the spacetime generated by bound states of Strominger-Vafa black holes holds strong resemblance to the Schwarzschild black hole. However, it differs significantly at a scale much larger than the horizon scale, particularly around the photon ring. Nevertheless, the bound states exhibit ultra-compact geometry and highlight the characteristics of novel physics emerging from the resolution of the Schwarzschild black hole into a novel spacetime structure with a clear microscopic origin in string theory, such as branes and antibranes.

\section{Microstate construction}

The previous construction facilitates a counting of states within a neutral brane/antibrane system involving both BPS and anti-BPS black holes in supergravity.  To construct explicit microstates, the extremal black holes must be resolved in terms of smooth, horizonless and topologically-nontrivial geometries. While not all microstates should admit such a classical description, it is possible to build coherent (though atypical) states in the phase space.

Large families of microstate geometries of the five-dimensional extremal black holes have been constructed to date \cite{Warner:2019jll,*Bena:2022ldq,*Bena:2022rna}. Two known categories are superstrata \cite{Bena:2015bea,*Bena:2017xbt,*Ceplak:2018pws,*Heidmann:2019zws,*Heidmann:2019xrd,*Shigemori:2020yuo} and multicenter bubbling geometries \cite{Bena:2007kg,*Heidmann:2017cxt,*Bena:2017fvm,*Bena:2018bbd,*Heidmann:2018vky} \footnote{Although additional generic microstate geometries with a larger phase space are anticipated, their supergravity descriptions remain incomplete \cite{Bena:2022wpl,*Bena:2022fzf,*Bena:2023rzm}.}.

Both five-dimensional black holes in our bound state can be locally resolved deep within their AdS$_2$ near-horizon throats by capping it off with two microstate geometries. The interaction between both microstates can be controlled by considering scaling geometries so that it will be mainly given by the black hole values.  Explicit bubbling geometries can be found numerically,  or potentially analytically by selecting the simplest geometries. 

\begin{figure}
\begin{center}
\includegraphics[width= 0.4 \textwidth]{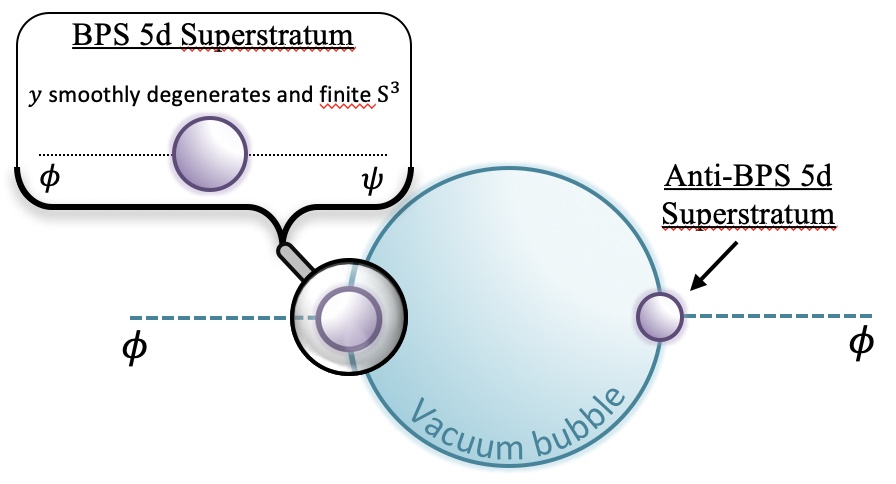}
\caption{\textit{Schematic description of a smooth, horizonless and neutral bound state of two superstrata held apart by a KK bubble.}}
\label{fig:BS2MG}
\end{center}
\vspace{-0.3cm}
\end{figure}

In Fig. \ref{fig:BS2MG}, we depict a typical smooth and horizonless geometry achievable by substituting both extremal black holes with one of their superstratum microstates. The resulting configuration corresponds to a bound state composed of two superstrata: one BPS and one anti-BPS, separated by the same smooth KK bubble as in the case of the black hole bound state. Apart from a small region near the poles of the bubble, the smooth solution is indistinguishable from the black hole bound state.

Furthermore, superstrata cover a significant fraction of the total phase space of the Strominger-Vafa black hole. Specifically, the entropy of the superstratum ensemble scales like $N_P^{1/4} \sqrt{N_1 N_5}$ \cite{Shigemori:2019orj,*Mayerson:2020acj}. By extending this result with \eqref{eq:MacroQuantCharges}, we anticipate the existence of $e^{\mathcal{M}^{5/3}}$ smooth, horizonless geometries corresponding to coherent microstates of the neutral bound state of two Strominger-Vafa black holes.  If it does not scale with the Schwarzschild entropy,  it significantly exceeds the entropy bound that ordinary matter can reach ($S < S_\text{Schw}^{3/4} \propto \mathcal{M}^{3/2}$) \cite{tHooft:1993dmi,*Frampton:2008mw}. Therefore, our approach establishes a strong connection between the microstructure of supersymmetric black holes and the microstructure of non-supersymmetric and neutral systems in string theory that have a phase space as vast as the entropy of a Schwarzschild black hole.

There is an other approach to build smooth horizonless geometries indistinguishable from the previous black hole bound state.  This consists in exploiting the generalized Ernst formalism in type IIB and in replacing the extremal black holes with near-extremal bubbles supported by the same amount of brane flux \cite{Bah:2022yji,Bah:2023ows}.  This resolution scheme differs from the first approach by not preserving the local extremality of the sources. Despite this, the advantage lies in achieving a fully analytical construction by using the Ernst formalism in higher dimensions, as demonstrated in other similar contexts \cite{Bah:2022yji,Bah:2023ows}.

\section{Outlook}
\label{sec:Disc}

The bound state presented in this article is a highly polarized configuration where the BPS and anti-BPS constituents are localized at two centers. One could consider more generic solutions where the branes and antibranes are distributed along the bubble using the solution of Ernst equations sourced by $N$ extremal centers \cite{HeidmannMehta}.  It would be interesting to investigate whether accounting for all bound states in supergravity yield an ensemble with an entropy of $S_\text{Schw}$.

A crucial direction for further study concerns stability. While the bound state is in gravitational equilibrium, it is likely unstable. First, it would be interesting to explore the classical stability of the solutions and the potential existence of Gregory-Laflamme modes \cite{Gregory:1994bj}, which could force the black holes to merge. Note that this merging cannot occur by having the black holes simply ``roll'' along the bubble, as such a process would require a change of topology that cannot be achieved by classical perturbation. Therefore, the black holes must remain at the poles of the bubble, but they can grow along the bubble until they fully annihilate it.

\noindent Second, it would be interesting to investigate thermodynamic instability, whereby the black holes might radiate and decay. This could occur through the separation of virtual pairs of branes and antibranes, similar to Hawking radiation, with the difference that this can happen at the bubble,  in the region between the black holes. This process would cause the bound state to lose energy by having the extremal black holes lose charges. An interesting aspect to explore is whether this decay can be associated with a form of Hawking radiation of the system at the effective temperature derived in equation \eqref{eq:FirstLaw}.

The fact that the bound state is classically indistinguishable from a Zipoy-Voorhees metric,  rather than the Schwarzschild black hole itself deserves further investigation.  We believe that similar constructions in type IIB involving near-extremal sources, rather than extremal ones, might solve this issue \cite{Dulac:2024cso}. Since near-extremal D1-D5-P black holes also have a well-defined microscopic description \cite{Callan:1996dv}, achieving such constructions would represent a significant breakthrough, potentially providing the first microscopic description of the Schwarzschild black hole using brane/antibrane bound states in supergravity.

Furthermore, in this study, we have focused on neutral bound states of Strominger-Vafa black holes, featuring an equal number of branes and antibranes, to facilitate comparison with a Schwarzschild black hole. However, we can explore scenarios with different numbers of branes and antibranes, resulting in a nonzero net charge. This investigation would allow us to compare such bound states with non-extremal Reissner-Nordström black holes and assess whether they offer a microscopic explanation for their entropy. The near-extremal limit is particularly intriguing as it would establish connections with previous works \cite{Callan:1996dv,Horowitz:1996ay}.

Despite the theoretical nature of the results presented in this paper, we can explore how the novel spacetime structure of branes and antibranes, replacing the Schwarzschild black hole, may manifest in new observables. This consists in deriving the gravitational signature of the solutions and studying their response to perturbations. One avenue of research could focus on the light scattering properties and imaging simulation of these geometries, comparing them to the Schwarzschild black hole, as done in previous studies \cite{Bacchini:2021fig,Heidmann:2022ehn,Sui:2023yay} for similar string-theoretical solutions. Another direction would involve analyzing the quasi-normal mode spectrum of the bound states and investigating how the internal structure modifies the Schwarzschild spectrum, as explored in previous works \cite{Bueno:2017hyj,Bena:2019azk,PhysRevD.106.024041,Maggio:2021ans,Heidmann:2023ojf}. Such investigations could shed light on whether the atypical string-theoretic structure constructed in this paper and that manifests below the Schwarzschild photon ring might impact gravitational wave signals.

\vspace{0.2cm}
\noindent
{\bf Acknowledgments}  I would like to thank  Ibrahima Bah, Iosif Bena, Bogdan Ganchev, Marcel Hughes,  Samir D. Mathur and Madhur Mehta for useful discussions,  especially Ibou and Iosif for providing valuable advice on organizing the results. The work is supported by the Department of Physics at The Ohio State University. 

\vspace{0cm}
\bibliographystyle{apsrev4-2}
\bibliography{microstates}

\end{document}